\documentclass[sigconf,balance=true,hidelinks,table]{acmart}

\usepackage{hyperref}
\usepackage{amsmath,amsfonts}
\usepackage{algorithmic}
\usepackage[ruled,linesnumbered,noend]{algorithm2e}
\usepackage{xcolor}
\usepackage{graphicx}
\usepackage{textcomp}
\usepackage{xcolor}
\usepackage{nccmath}
\usepackage{environ}
\usepackage{tcolorbox}
\usepackage{soul}
\usepackage{booktabs}
\usepackage{multirow}
\usepackage{array}
\usepackage{float}
\usepackage{makecell}
\usepackage{flushend}
\usepackage{enumitem}
\usepackage{hhline}
\usepackage{comment}
\usepackage{epstopdf} 
\usepackage{caption}

\usepackage{xcolor}
\usepackage{listings}

\usepackage{scalerel}

\NewEnviron{myequation}{%
\begin{equation}
\scalebox{1}{$\BODY$}
\end{equation}
}

\newcommand{\fig}{Fig.}
\newcommand{\tab}{Table}

\newcommand{\CodeBERT}{CodeBERT}

\newcolumntype{P}[1]{>{\centering\arraybackslash}p{#1}}
\newcolumntype{R}[1]{>{\raggedleft\arraybackslash}p{#1}}

\newcommand{\code}[1]{\textnormal{\texttt{#1}}}

\usepackage{tikz}
\usetikzlibrary{shapes.arrows}

\definecolor{brightgreen}{rgb}{0.4, 1.0, 0.0}
\definecolor{ferrarired}{rgb}{1.0, 0.11, 0.0}

\acmConference[Secure Software 2024]{The 4th International Workshop on Software Security: Challenges, Opportunities, and Lessons Learned}{18–21 June, 2024}{Salerno, Italy}

\copyrightyear{2024}
\acmYear{2024}
\setcopyright{rightsretained}
\acmConference[EASE 2024]{28th International Conference on Evaluation and Assessment in Software Engineering}{June 18--21, 2024}{Salerno, Italy}
\acmBooktitle{28th International Conference on Evaluation and Assessment in Software Engineering (EASE 2024), June 18--21, 2024, Salerno, Italy}\acmDOI{10.1145/3661167.3661281}
\acmISBN{979-8-4007-1701-7/24/06}

\begin{document}

\title{Software Vulnerability Prediction in Low-Resource Languages: An Empirical Study of CodeBERT and ChatGPT}

\author{Triet Huynh Minh Le}
\affiliation{\institution{CREST - The Centre for Research on Engineering Software Technologies, The University of Adelaide}
\city{Adelaide}
\country{Australia}}
\affiliation{\institution{Cyber Security Cooperative Research Centre, Australia}
\city{}
\country{}}
\email{triet.h.le@adelaide.edu.au}

\author{M. Ali Babar}
\affiliation{\institution{CREST - The Centre for Research on Engineering Software Technologies, The University of Adelaide}
\city{Adelaide}
\country{Australia}}
\affiliation{\institution{Cyber Security Cooperative Research Centre, Australia}
\city{}
\country{}}
\email{ali.babar@adelaide.edu.au}

\author{Tung Hoang Thai}
\affiliation{\institution{CREST - The Centre for Research on Engineering Software Technologies, The University of Adelaide}
\city{Adelaide}
\country{Australia}}
\email{hoangtung.thai@adelaide.edu.au}

\begin{abstract}

\textbf{Background}: Software Vulnerability (SV) prediction in emerging languages is increasingly important to ensure software security in modern systems. However, these languages usually have limited SV data for developing high-performing prediction models.
\textbf{Aims}: We conduct an empirical study to evaluate the impact of SV data scarcity in emerging languages on the state-of-the-art SV prediction model and investigate potential solutions to enhance the performance.
\textbf{Method}: We train and test the state-of-the-art model based on CodeBERT with and without data sampling techniques for function-level and line-level SV prediction in three low-resource languages -- Kotlin, Swift, and Rust. We also assess the effectiveness of ChatGPT for low-resource SV prediction given its recent success in other domains.
\textbf{Results}: Compared to the original work in C/C++ with large data, CodeBERT's performance of function-level and line-level SV prediction significantly declines in low-resource languages, signifying the negative impact of data scarcity. Regarding remediation, data sampling techniques fail to improve CodeBERT; whereas, ChatGPT showcases promising results, substantially enhancing predictive performance by up to 34.4\% for the function level and up to 53.5\% for the line level.
\textbf{Conclusion}: We have highlighted the challenge and made the first promising step for low-resource SV prediction, paving the way for future research in this direction.

\end{abstract}

\ccsdesc[500]{Security and privacy~Software security engineering}

\keywords{Software vulnerability, Software security, Large language models, ChatGPT, Empirical study}

\maketitle

\section{Introduction}

Software Vulnerabilities (SVs) present tremendous threats to the security and dependability of software systems. Given the growing scale and complexity of software applications~\cite{le2019automated}, there is a pressing requirement for the automatic detection of SVs~\cite{hanif2021rise}.
There has been a growing use of Deep Learning models for SV detection, particularly for identifying potentially vulnerable functions and lines~\cite{le2020deep,lin2020software,le2022survey,fu2022linevul}.
Among these models, CodeBERT has been demonstrated to be the state-of-the-art for SV prediction~\cite{fu2022linevul,steenhoek2023empirical}.
The successful development of these SV prediction models heavily depends on the availability of SV datasets~\cite{croft2022data}.

Datasets necessary for constructing models predicting SVs are often lacking for emerging (recent yet widely used) programming languages. We refer to this scenario as ``\textit{low-resource SV prediction}.'' Our investigation of three emerging languages, namely Kotlin, Swift, and Rust, has revealed their SV data is merely 0.2\% to 0.8\% the size of that for C/C++, the extensively studied language in the literature. The state-of-the-art SV prediction model, utilizing CodeBERT~\cite{fu2022linevul}, excels at SV prediction for C/C++ with over 90\% F1-Score.
However, its performance for emerging languages with the demonstrated limited data is likely to be affected, yet the extent of the impact remains unknown.
To tackle the data scarcity, besides traditional data sampling techniques, ChatGPT has shown exceptional performance across tasks~\cite{fan2023large}, including low-resource contexts. Nevertheless, to the best of our knowledge, its applicability to SV prediction in low-resource languages has not been explored.

To answer these questions, we conduct an empirical study on the performance of predicting SVs in three emerging yet low-resource languages, namely Kotlin, Swift, and Rust. We first evaluate the performance of the state-of-the-art CodeBERT based model for the tasks.
We then investigate whether data sampling techniques such as random over-sampling and random under-sampling, aiming at tackling data scarcity, can improve the performance of CodeBERT. We also explore the potential use of ChatGPT with few-shot learning and fine-tuning for low-resource SV prediction. Our findings are expected to provide evidence-based knowledge about the extent to which we can reuse the state-of-the-art SV model for emerging languages with limited data and whether data sampling or ChatGPT can improve the performance in this practical scenario.

Our key \textbf{contributions} can be summarized as follows:

\begin{itemize}[noitemsep,topsep=0pt,leftmargin=*]
    \item We are the first to automate function-level and line-level SV prediction in low-resource languages, i.e., Kotlin, Swift, and Rust.
    \item We empirically demonstrate the performance of function-level and line-level SV prediction in low-resource languages. Compared to C/C++ with abundant data on which the state-of-the-art CodeBERT model was originally trained, the model obtains much lower performance of SV prediction in low-resource languages (e.g., 0.35 vs. 0.9 for the function level). We also show that data sampling techniques cannot improve the performance. On the other hand, ChatGPT enhances the performance by up to 34.4\% for the function level and 53.5\% for the line level. Overall, the performance gaps between low-resource and abundant-resource languages are still large, motivating further research.
    \item We share our data and code for future research at~\cite{reproduction_package_ease2024}.
\end{itemize}

\noindent \textbf{Paper structure}. Section~\mbox{\ref{sec:background}} introduces the related work and motivation for the study. Section~\mbox{\ref{sec:rqs}} presents the research questions.
Section~\ref{sec:setup} describes the methods used to answer the questions.
Section~\mbox{\ref{sec:results}} reports the results to each of the questions.
Section~\mbox{\ref{sec:threats_to_validity}} discusses the threats to validity. Section~\mbox{\ref{sec:conclusions}} concludes the study.

\section{Background and Motivation}\label{sec:background}

\subsection{Software Vulnerability (SV) Prediction}

In recent years, data-driven approaches like Machine Learning and Deep Learning models have been widely used to automate the identification/prediction of SVs in source code (e.g.,~\cite{lin2020software,hanif2021rise,le2021deepcva}). The predictions have been performed on various levels of granularity, ranging from package/file to function and line. The more fine-grained function and statement levels can reduce inspection effort for developers~\cite{hin2022linevd,fu2022linevul}. Thus, the two aforementioned fine granularities have become the standard for SV prediction, and thus they are also adopted for our investigations.

\begin{figure}[t]
    \centering
    \includegraphics[trim={12cm 1cm 12cm 1cm},clip,width=\columnwidth,keepaspectratio]{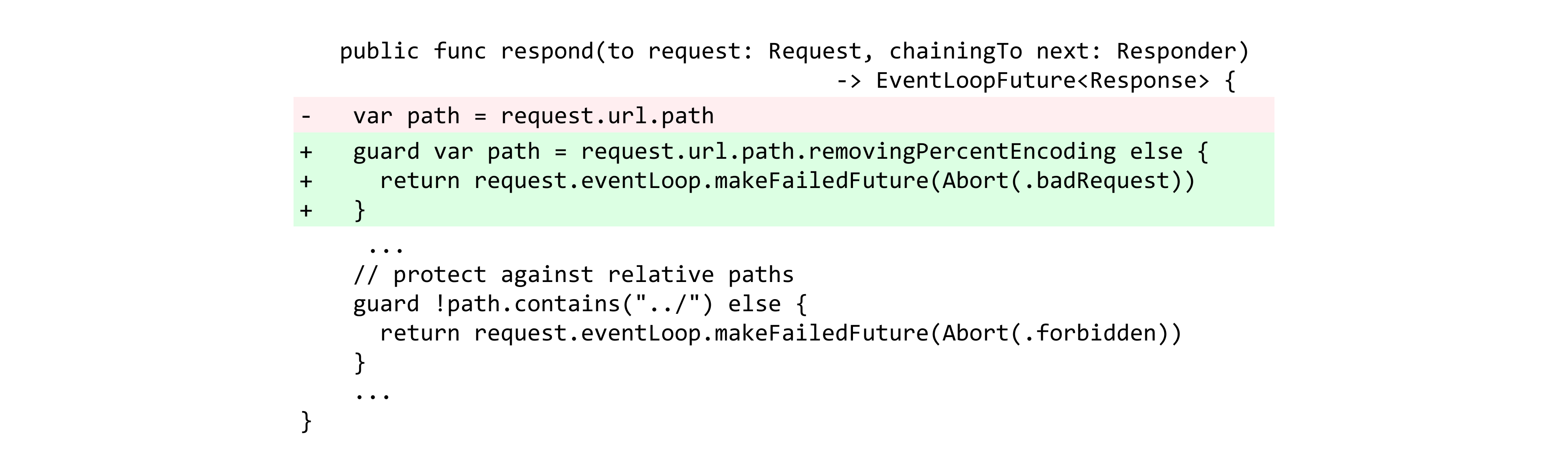}
    \caption{Exemplary vulnerable function and lines corresponding to \textit{CVE-2020-15230} extracted from the respective vulnerability-fixing commit in the \textit{vapor} project in Swift.}
    \label{fig:vuln_ex}
\end{figure}

\fig~\ref{fig:vuln_ex} gives an example of an SV (CVE-2020-15230) in the \textit{vapor} project written in Swift. This SV originates from the lines ``\code{var path = request.url.path}'' in the \code{respond} function. This line directly assigns user's input to the \code{path} variable without performing any sanitization, and this variable is then checked for relative paths on the line ``\code{guard !path.contains("../") else \{}''. However, attackers can bypass this check by replacing the dot (``.'') with the percent symbol (\code{\%2E}) in the variable, potentially leading to a path traversal SV. This SV was fixed in the commit \textit{cf1651f} in which any percent symbol would be removed from the \code{path} variable to ensure that all the subsequent checks would be properly performed.

\subsection{Challenges of SV Prediction in Low-Resource Languages}

The increasing demand of the software industry has given birth to a wide range of new programming languages. Many of these newly introduced languages have later become widely used for software development because of their unique features and advantages, such as Kotlin, Swift, and Rust.
For example, Kotlin finds extensive applications in Android mobile development; Swift has become the language of choice for iOS and macOS applications, emphasizing safety and performance; Rust, known for its memory safety and low-level control, is increasingly used for systems programming, particularly in security-critical contexts.

While emerging languages play pivotal roles in modern software development, the amount of data, especially concerning SVs, is much more limited compared to traditional languages like C/C++.
This argument has been strongly supported by our analysis of SV data in these languages (see \tab~\ref{tab:dataset-size}). Specifically, we found that the numbers of SVs in Kotlin, Swift, and Rust were 1,598, 389, and 157 times smaller than that of more established languages like C/C++, respectively.
The demonstrated scarcity of SV data can significantly hamper the performance of downstream data-driven SV prediction models for these languages given the data hungriness of these models~\cite{croft2022data}.
Thus, our study is the first to evaluate the performance of CodeBERT~\cite{fu2022linevul}, the state-of-the-art SV prediction model, in such low-resource languages. Additionally, we aim to explore the feasibility of using ChatGPT to address the data scarcity challenge in these emerging languages, given ChatGPT's success in many other low-resource Software Engineering tasks~\cite{fan2023large}.

\subsection{Large Language Models for SV Management}

The literature witnesses increasing attention and use of large language models, especially ChatGPT, for SV prediction~\cite{zhou2024large}. Cheshkov et al.~\cite{cheshkov2023evaluation} leveraged ChatGPT for identifying SVs of five different types (CWE-IDs). Zhang et al.~\cite{zhang2023prompt} improved the performance of the task by leveraging prompt engineering with ChatGPT.
Pearce et al.~\cite{pearce2023examining} assessed the performance of various large language models including ChatGPT for SV fixing in the zero-shot scenario.
In an attempt to automate various tasks for SV management, Fu et al.~\cite{fu2023chatgpt} investigated ChatGPT with prompt engineering for SV classification, severity assessment, and fixing.
Overall, these studies have shown promising results of ChatGPT for SV analysis tasks, especially when using few-shot learning with prompt engineering.
Fundamentally, our work is different from the current literature as we focus on empirically evaluating ChatGPT for SV detection in \textit{low-resource languages}, which is an important and practical problem in modern software development. We are also the first to investigate fine-tuning ChatGPT besides prompt engineering for function-level and line-level SV prediction in low-resource languages.

\section{Research Questions}
\label{sec:rqs}

We answer the following Research Questions (RQs) to investigate the performance of SV prediction in low-resource languages.

\begin{itemize}
    \item \textbf{RQ1}: How well does the state-of-the-art CodeBERT based model detect SVs in low-resource languages?
    \item \textbf{RQ2}: Can ChatGPT improve the performance for low-resource SV prediction?
\end{itemize}

\section{Case Study Setup}
\label{sec:setup}

This section describes the datasets and the models, i.e., CodeBERT and ChatGPT, used for low-resource SV prediction as well as the evaluation procedure for these models.

\subsection{Datasets}

We leveraged the methods and tools provided by CVEfixes~\cite{bhandari2021cvefixes} to curate SV data, including vulnerable functions and lines, for low-resource languages.
Essentially, the data collection starts with SV-fixing commits. In these commits, the functions encompassing lines changed are considered vulnerable; otherwise, they are non-vulnerable. The deleted lines are labeled as vulnerable lines. This data curation process follows the same practice of Big-Vul~\cite{fan2020ac}, the largest SV dataset in C/C++ widely used in the literature.

Regarding low-resource SV prediction, we selected three languages, Kotlin, Swift, and Rust, for two reasons.
Firstly, these three languages have an extremely limited number (< 100) of vulnerable functions, making them directly relevant to our focus on low-resource SV prediction.
Secondly, these languages, with the first release recently from 2014 to 2016, are being extensively used in practice, as evidenced by the Stack Overflow survey in 2023.\footnote{https://survey.stackoverflow.co/2023/\#most-popular-technologies-language-prof}
The statistics of the data collected for each language are given in \tab~\ref{tab:dataset-size}.

\begin{table}[t]
\fontsize{7.5}{8.5}\selectfont
\centering
\caption{Data statistics in Kotlin, Swift, and Rust languages.}
\begin{tabular}{l|lll}
\hline
\textbf{Statistic}                  & \textbf{Kotlin} & \textbf{Swift} & \textbf{Rust} \\ \hline
Distinct Projects                 & 4               & 7              & 19            \\ \hline\hline
Vulnerable functions    & 20              & 36             & 90            \\ \hline
Non-vulnerable functions & 98              & 449            & 1,109          \\ \hline\hline
Vulnerable lines         & 45              & 104            & 350           \\ \hline
Non-vulnerable lines      & 1,208            & 8,091           & 26,479         \\ \hline
\end{tabular}%
\label{tab:dataset-size}
\end{table}

\subsection{SOTA for SV prediction with CodeBERT}

The fine-tuned  CodeBERT~\cite{feng2020codebert} model by Fu et al.~\cite{fu2022linevul} currently stands as the State-Of-The-Art (SOTA) for function-level and line-level SV prediction~\cite{steenhoek2023empirical}; thus, it was employed for our investigations. The model derives code representations capturing both syntactic and semantic information.
Function-level predictions are crafted using a Transformer-based architecture; the most vulnerable lines within these functions are then pinpointed using attention scores from the trained Transformer model.
Following Fu et al.~\cite{fu2022linevul}, we also fine-tuned CodeBERT for each of the three studied languages to predict vulnerable functions and lines.
The hyperparameters of \CodeBERT{} were adapted from previous studies (e.g.,~\cite{steenhoek2023empirical,croft2023data}) as follows: \textit{epochs}: 10, \textit{learning rate}: 1e-5, and \textit{feature embedding size}: 768. Despite exploring alternative values, no significant performance improvement was observed.
To tackle data scarcity, we also applied Random Over-Sampling (ROS) and Random Under-Sampling (RUS) to \textit{only} the training sets before fine-tuning CodeBERT.

\subsection{ChatGPT for SV prediction}

We utilized the APIs of ChatGPT based on GPT-3.5-Turbo~\cite{gpt_turbo} to develop prediction models for identifying SVs in low-resource languages. Our approach, called prompt chaining, decomposed the task into two subtasks: first detecting vulnerable functions and then predicting vulnerable lines based on the result of the first step.

\subsubsection{Function-level prediction}

We explored two techniques for function-level SV prediction: few-shot learning and fine-tuning.

\begin{figure}[t]
    \centering
    \includegraphics[width=\columnwidth,keepaspectratio]{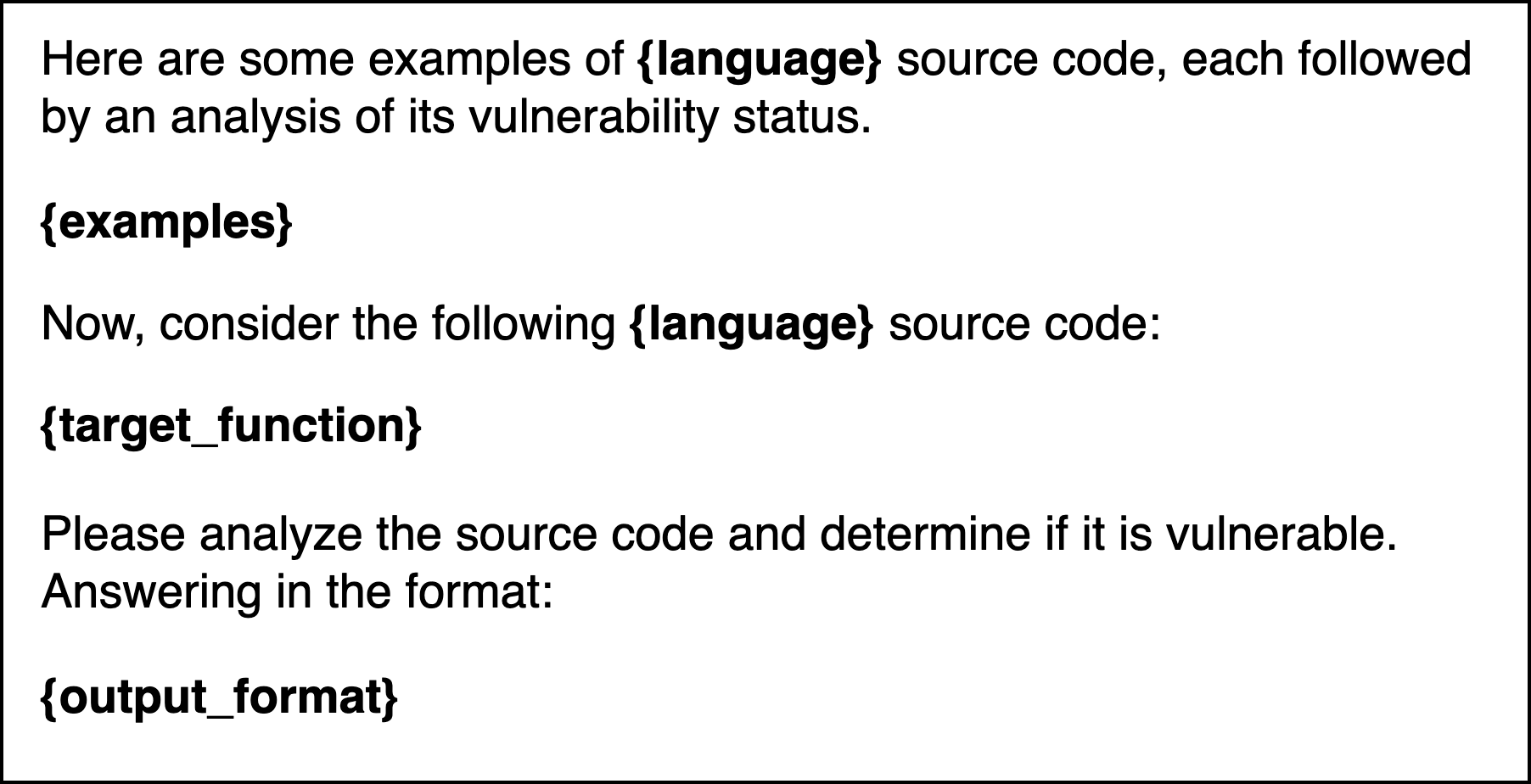}
    \caption{Prompt to use ChatGPT with few-shot learning for function-level SV prediction.}
    \label{fig:few-shot learning prompt}
\end{figure}

\noindent \textbf{Few-shot learning}.
We crafted a specialized prompt that directed the model to identify SVs within a specified function (see \fig~\ref{fig:few-shot learning prompt}). We incorporated examples showcasing vulnerable and non-vulnerable functions to enhance the model's generalizability across various instances. This method leverages the inherent capability of ChatGPT from vast knowledge to learn from a few SV examples and then draw inferences about new SVs from the contextual clues provided within the given samples.

The prompt of the few-shot learning approach includes:

\begin{enumerate}
    \item \textbf{language:} the programming language of examples and the target function.
    \item \textbf{examples:} the list of example functions. Each function is formatted as \texttt{<input, output>}.
    \begin{itemize}
        \item input: the code of the function.
        \item output: the vulnerability status of the function code.
    \end{itemize}
    \item \textbf{target\_function:} The function we need to predict SV for.
    \item \textbf{output\_format:} The expected format of the model's response. It is set to ``vulnerable'' or ``non-vulnerable''.
\end{enumerate}

We carefully designed the few-shot learning prompt, balancing performance with data efficiency and GPT-3.5-turbo model limitation. Our experiments revealed that 10 examples, including nine vulnerable and one non-vulnerable function, achieved this optimal balance. This choice was driven by the limited availability of labeled vulnerable code in training datasets and the maximum token limit of the GPT-3.5-turbo model. Including longer prompts with more examples would exceed this limit, hindering the evaluation stage. Additionally, we found that including too many non-vulnerable examples in the training set also hindered model performance. Conversely, using nine vulnerable examples alongside a single non-vulnerable example demonstrably improved the model's ability to generalize and achieve better performance.

\begin{figure}[t]
    \centering
    \includegraphics[width=\columnwidth,keepaspectratio]{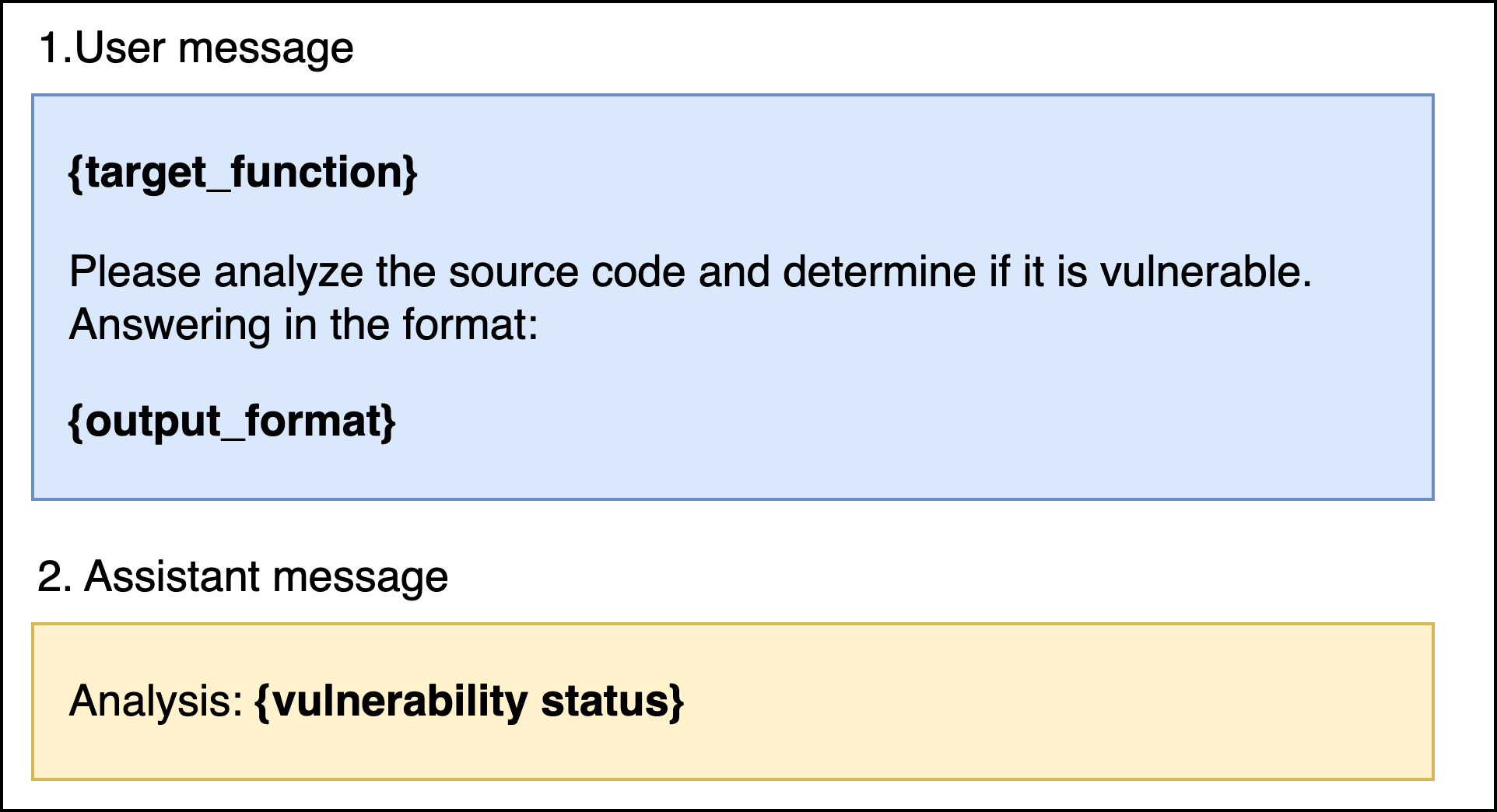}
    \caption{Prompt to use ChatGPT with fine-tuning for function-level SV prediction.}
    \label{fig:fine-tuning prompt}
\end{figure}

\noindent \textbf{Fine-tuning}.
Fine-tuning can address the limitations of the few-shot learning approach, particularly the constraints on the number of examples that can be incorporated within a single prompt illustrated in \fig~\ref{fig:fine-tuning prompt}.
Specifically, fine-tuning uses our labeled training data to change the weights of GPT-3.5-Turbo to make the model become more specialized in SV prediction.

\begin{figure}[t]
    \centering
    \includegraphics[width=\columnwidth,keepaspectratio]{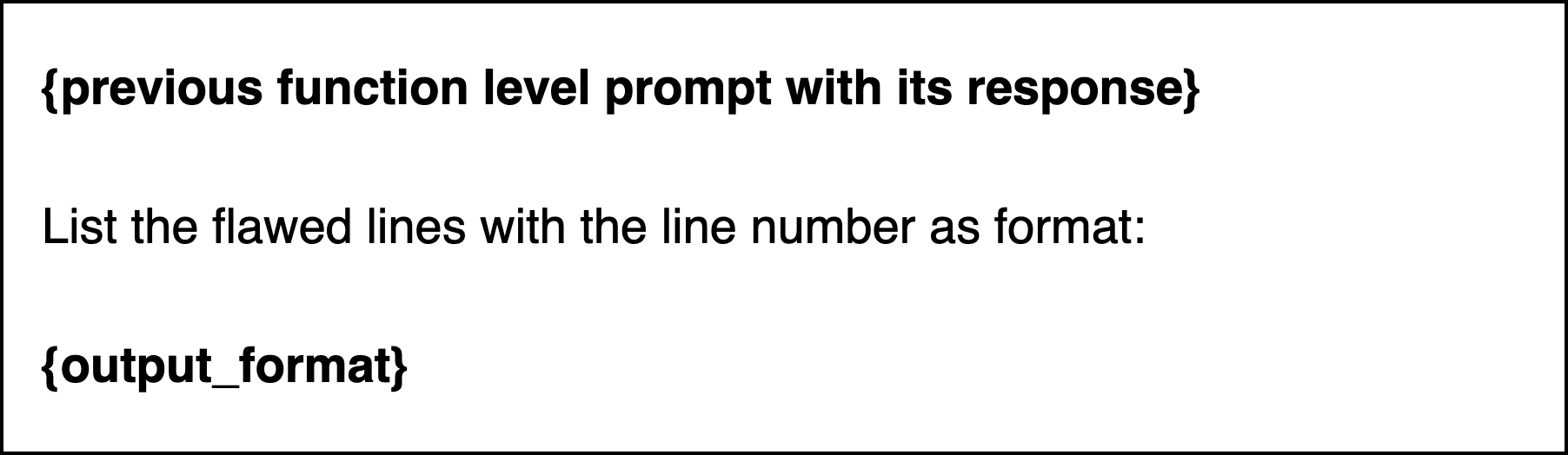}
    \caption{Prompt to use the ChatGPT model trained at the function level for line-level SV prediction.}
    \label{fig:line-level prompt}
\end{figure}

Based on the OpenAI's API, the fine-tuning prompt includes:

\begin{enumerate}
    \item \textbf{User Message:} This element serves as input data for fine-tuning. It includes the \textit{target\_function} parameter that specifies the code of each function in a training set and the \textit{output\_format} parameter including the expected response, i.e., vulnerable or non-vulnerable.
    \item \textbf{Assistant Message:} This element includes the ground truth of the prediction, explicitly indicating the training code snippet's vulnerability status (``vulnerable'' or ``non-vulnerable''). This clear distinction guides the model in learning the desired output for each training instance.
\end{enumerate}

After fine-tuning, the model can handle zero-shot learning, allowing the model to predict SVs without requiring additional examples within the prompt itself. This streamlines the process and significantly improves response times. Similar to CodeBERT in RQ1, we also investigated fine-tuning ChatGPT with random over-sampling and random under-sampling. We found that over-sampling worked best with ChatGPT, and thus, we would report the ChatGPT's results based on this setting.

\subsubsection{Line-level prediction}

If a function was predicted as vulnerable by ChatGPT with either few-shot learning or fine-tuning, we would leverage a line-level prediction prompt, given in ~\fig~\ref{fig:line-level prompt}, to incorporate both the initial prediction and its output as context.
We designed this prompt based on the intuition from CodeBERT~\cite{fu2022linevul} that the model would know the vulnerable lines, a.k.a the reasons, contributing to its function-level SV prediction. This assumption would be likely valid if the model already correctly predicted the vulnerable functions.
This comprehensive view allows the model to focus its attention on specific lines within the function that are most likely vulnerable.

\subsection{\textbf{Model Evaluation}}
\label{subsubsec:model_evaluation}

\noindent \textbf{Evaluation technique}. We used a 10-round evaluation for the models. Each round split the vulnerable and non-vulnerable functions into training, validation, and testing sets at ratios of 60:20:20, respectively. This ensured sufficient samples for testing function-level and line-level SV prediction. To prevent data leakage, we excluded all duplicate training entries/functions from the validation and testing sets in each round.
For CodeBERT, we employed the early stopping strategy~\mbox{\cite{hoang2019deepjit}}, i.e.,  stopping training if the validation performance did not enhance in the last five epochs within a round. We selected the optimal configurations for CodeBERT and ChatGPT based on the highest performance averaging all validation sets.

\noindent \textbf{Evaluation measures}. We used established evaluation measures for function-level and line-level SV prediction in our investigations. For function-level SV prediction, we utilized \textit{F1-Score}, \textit{Precision}, and \textit{Recall}, common measures widely applied in prior SV prediction studies (e.g.,~\cite{zhou2019devign,fu2022linevul,le2022use,steenhoek2023empirical}). F1-Score, being the harmonic mean of Precision and Recall, was chosen for optimal model selection. Reported results reflected the average performance on all the \textit{testing} sets of these optimal models that were determined by the highest validation F1-Score.

For line-level SV prediction, we calculated \textit{Top-3 Accuracy} and \textit{Initial False Alarm} (IFA), standard measures for assessing model interpretability~\cite{li2021vulnerability,fu2022linevul}.\footnote{We did not use Top-10 Accuracy, as in CodeBERT, as the average LOC of Kotlin functions was around 10, i.e., vulnerable lines almost always rank in the top list.} These measures gauge the effectiveness of localizing vulnerable lines, especially the first one, aiding developers in initiating inspection. We also used \textit{Effort@20\%Recall} and \textit{Recall@1\%LOC}~\cite{fu2022linevul} to assess performance while considering the effort developers would need to inspect vulnerable lines.

\section{Experimental Results}
\label{sec:results}

\begin{table}[t]
\fontsize{7.5}{8.5}\selectfont
\centering
\caption{Function-level SV prediction performance of CodeBERT in low-resource languages. \textbf{Notes}: RUS is Random Under-Sampling; ROS is Random Over-Sampling. The base model of CodeBERT did not use either ROS or RUS.}
\begin{tabular}{lcccc}
\hline
\textbf{Lang.}        & \textbf{Model} & \textbf{F1-Score} & \textbf{Precision} & \textbf{Recall} \\ \hline
\multirow{3}{*}{Kotlin}  & CodeBERT (Base) & 0.25              & \textbf{0.30}      & 0.22            \\
                         & CodeBERT (RUS)  & \textbf{0.32}     & 0.24               & \textbf{0.47}   \\
                         & CodeBERT (ROS)  & \textbf{0.32}              & 0.29               & 0.37            \\ \hline
\multirow{3}{*}{Swift}   & CodeBERT (Base) & \textbf{0.37}     & \textbf{0.44}      & 0.33            \\
                         & CodeBERT (RUS)  & 0.34              & 0.24               & \textbf{0.63}   \\
                         & CodeBERT (ROS)  & 0.31              & 0.25               & 0.42            \\ \hline
\multirow{3}{*}{Rust}    & CodeBERT (Base) & \textbf{0.43}              & 0.44               & 0.43            \\
                         & CodeBERT (RUS)  & 0.27              & 0.17               & \textbf{0.66}   \\
                         & CodeBERT (ROS)  & 0.42     & \textbf{0.46}      & 0.40            \\ \hline
                         \hline
\multirow{3}{*}{Avg.} & CodeBERT (Base) & \textbf{0.35}     & \textbf{0.39}      & 0.33            \\
                         & CodeBERT (RUS)  & 0.31              & 0.22               & \textbf{0.59}   \\
                         & CodeBERT (ROS)  & \textbf{0.35}              & 0.33               & 0.40            \\ \hline
\end{tabular}%
\label{tab:function-level-results}
\end{table}

\subsection{RQ1: SOTA Model (CodeBERT) for SV Prediction in Low-Resource Languages}
\label{subsec:rq1_results}

The performance of the fine-tuned CodeBERT for predicting vulnerable functions in low-resource languages was still limited for all three studied languages (see \tab~\ref{tab:function-level-results}).
The base CodeBERT model attained 0.25 to 0.43 F1-Score for Kotlin, Swift, and Rust, respectively, compared to 0.91 F1-Score originally reported for C/C++~\cite{fu2022linevul}.
This finding supports our hypothesis earlier on that low-resource languages with much smaller sizes of data than C/C++ (i.e., 188,636 C/C++ functions, 10,900 of which are vulnerable) are likely to suffer from significant performance degradation.
It is also worth noting that CodeBERT's F1-Score rose as the dataset size increased, from Kotlin (the smallest size) to Rust (the largest size). We also observed similar increasing trends for Precision and Recall.

\begin{table}[t]
\fontsize{7.5}{8.5}\selectfont
\centering
\caption{Line-level SV prediction performance of CodeBERT in low-resource languages. \textbf{Note}: For the metric value, the higher is the better, except for IFA and Effort@20\%Recall.}
\begin{tabular}{lccccc}
\hline
\textbf{Lang.}        & \textbf{Model} & \makecell[c]{\textbf{Top-3}\\ \textbf{Accuracy}} & \textbf{IFA}  & \makecell[c]{\textbf{Effort@}\\ \textbf{20\%Recall}} & \makecell[c]{\textbf{Recall@}\\ \textbf{1\%LOC}} \\ \hline
\multirow{3}{*}{Kotlin}  & CodeBERT (Base) & \textbf{100\%}                                                    & \textbf{0.00}    & \textbf{0.013}                                                       & 0.111                                                            \\
                         & CodeBERT (RUS)  & 87.5\%                                                           & 1.38          & 0.023                                                                & 0.134                                                            \\
                         & CodeBERT (ROS)  & \textbf{100\%}                                                    & \textbf{0.00}    & \textbf{0.013}                                                       & \textbf{0.139}                                                   \\ \hline
\multirow{3}{*}{Swift}   & CodeBERT (Base) & 23.3\%                                                           & \textbf{6.70} & 0.032                                                                & 0.117                                                            \\
                         & CodeBERT (RUS)  & \textbf{31.0\%}                                                     & 9.25          & \textbf{0.024}                                                       & \textbf{0.250}                                                   \\
                         & CodeBERT (ROS)  & 21.7\%                                                           & 9.85          & \textbf{0.024}                                                       & 0.163                                                            \\ \hline
\multirow{3}{*}{Rust}    & CodeBERT (Base) & \textbf{58.7\%}                                                  & \textbf{7.69} & 0.027                                                                & 0.216                                                            \\
                         & CodeBERT (RUS)  & 43.9\%                                                           & 10.5         & 0.036                                                                & \textbf{0.306}                                                   \\
                         & CodeBERT (ROS)  & 40.2\%                                                           & 8.47          & \textbf{0.021}                                                       & 0.228                                                            \\ \hline
                         \hline
\multirow{3}{*}{Avg.} & CodeBERT (Base) & \textbf{60.7\%}                                                  & \textbf{4.80} & 0.024                                                                & 0.148                                                            \\
                         & CodeBERT (RUS)  & 54.1\%                                                           & 7.03          & 0.028                                                                & \textbf{0.230}                                                   \\
                         & CodeBERT (ROS)  & 53.9\%                                                           & 6.11          & \textbf{0.020}                                                       & 0.176                                                            \\ \hline
\end{tabular}%
\label{tab:line-level-results}
\end{table}

The performance of the fine-tuned CodeBERT model at line-level prediction varies more than the function level (see Table \ref{tab:line-level-results}). Unlike the relatively consistent performance of function-level prediction, the line-level prediction exhibited a fluctuating pattern across the Kotlin, Swift, and Rust languages.
Generally, we found a significant decrease in performance from Kotlin to Swift, followed by a slight improvement in Rust, suggesting that CodeBERT's performance in pinpointing vulnerable lines may be influenced by other factors beyond the dataset size. For example, a model might have learned code patterns for predicting vulnerable functions that may not align with the actual lines to be fixed by developers~\cite{steenhoek2023empirical}. The values of IFA, Effort@20\%Recall, and Recall@1\%LOC were also lower than those reported for C/C++~\cite{fu2022linevul}.

The results in Table \ref{tab:function-level-results} show that the data sampling techniques, i.e., ROS and RUS, did not consistently improve performance across all datasets.
For the function-level predictions, ROS tended to perform better than RUS and on par with the base model, while for the line-level predictions, ROS and RUS outperformed the base model by only one of four measures.
This inconsistency highlights the significant challenge of tackling the data scarcity issue when performing SV prediction in low-resource languages.
These results also motivate a need to explore alternative models for the tasks.

\begin{tcolorbox}
\textbf{RQ1 Summary}.
The performance of CodeBERT for low-resource SV prediction is still limited, as compared to C/C++ with large-sized data. Function-level prediction positively correlates with the data size.
Line-level prediction varies more across languages and is not directly affected by the dataset size. Data sampling techniques do not significantly improve function-level and line-level predictions.
\end{tcolorbox}

\begin{table}[t]
\fontsize{7.5}{8.5}\selectfont
\centering
\caption{Comparisons between ChatGPT and CodeBERT for function-level SV prediction in low-resource languages.}
\begin{tabular}{lcccc}
\hline
\textbf{Lang.}        & \textbf{Model}  & \textbf{F1-Score} & \textbf{Precision} & \textbf{Recall} \\ \hline
\multirow{3}{*}{Kotlin}  & GPT Fine-tuning & 0.34              & 0.37               & 0.32            \\
                         & GPT Few-shot    & \textbf{0.43}     & \textbf{0.43}      & 0.44            \\
                         & CodeBERT (Best)  & 0.32              & 0.24               & \textbf{0.47}   \\ \hline
\multirow{3}{*}{Swift}   & GPT Fine-tuning & \textbf{0.40}              & \textbf{0.55}      & 0.32            \\
                         & GPT Few-shot    & 0.34              & 0.36               & 0.32            \\
                         & CodeBERT (Best)  & 0.37     & 0.44               & \textbf{0.33}   \\ \hline
\multirow{3}{*}{Rust}    & GPT Fine-tuning & \textbf{0.44}     & \textbf{0.49}      & 0.40            \\
                         & GPT Few-shot    & 0.09              & 0.09               & 0.09            \\
                         & CodeBERT (Best)  & 0.43     & 0.44               & \textbf{0.43}   \\ \hline\hline
\multirow{3}{*}{Avg.} & GPT Fine-tuning & \textbf{0.40}     & \textbf{0.47}      & 0.35            \\
                         & GPT Few-shot    & 0.29              & 0.29               & 0.28            \\
                         & CodeBERT (Best)  & 0.37              & 0.38               & \textbf{0.40}   \\ \hline
\end{tabular}%
\label{tab:gpt-results-function-level}
\end{table}

\subsection{RQ2: ChatGPT for SV Prediction in Low-Resource Languages}
\label{subsec:rq2_results}

Our investigations into ChatGPT with few-shot learning and fine-tuning as alternative models to CodeBERT for SV prediction in low-resource datasets yielded promising results.
Note that we reported the results of fine-tuning ChatGPT with random over-sampling because it proved to be the most effective approach for this model type. For CodeBERT, we used the results of the best model obtained from RQ1 for each language.

\noindent \textbf{Function-level prediction}.
As shown in Table \ref{tab:gpt-results-function-level}, ChatGPT models, on average, produced 2.3\% to 34.4\% higher F1-Score than CodeBERT across all three datasets (Kotlin, Swift, Rust).
Notably, ChatGPT with few-shot learning demonstrated the best performance in the smaller Kotlin dataset. However, its effectiveness tended to diminish with larger and potentially more complex datasets like Rust. In contrast, ChatGPT with fine-tuning exhibited a higher F1-Score than CodeBERT consistently across all languages. Such stability of F1-Score suggests the robustness and scalability of ChatGPT with fine-tuning in handling datasets of varying sizes and complexities. The better overall performance (F1-Score) can be attributed to the higher Precision of ChatGPT than that of CodeBERT, meaning fewer false positives.
It is important to note that on average, the higher F1-Score and Precision values of ChatGPT over CodeBERT were statistically significant, based on the Wilcoxon signed-rank tests~\cite{wilcoxon1992individual} with $p$-values $<$ 0.01 and non-negligible effect sizes.\footnote{Effect size $(r) = Z / \sqrt{N}$; $Z$ is the $Z$-score statistic of the test and $N$ is sample size. When $r \geq 0.1$, the effect size is non-negligible~\cite{tomczak2014need}.}
Despite the improvements, the best F1 of ChatGPT is still much lower than that (0.91) of C/C++ with abundant SV data.

\noindent \textbf{Line-level SV prediction}.
For line-level prediction, ChatGPT with few-shot learning and fine-tuning outperformed CodeBERT across all the measures, except Recall@1\%LOC, as given in \tab~\ref{tab:gpt-results-line-level}. On average, the improvements of the ChatGPT variants over CodeBERT were 9.3\%, 53.5\%, and 50\%, for Top-3 Accuracy, IFA, and Effort@20\%Recall, respectively. There was no clear winner between the few-shot learning and fine-tuning variants of ChatGPT.
The line-level improvements of ChatGPT variants over the CodeBERT models were confirmed statistically significant, based on the Wilcoxon signed-rank tests~\mbox{\cite{wilcoxon1992individual}} with $p$-values $<$ 0.01 and non-negligible effect sizes.
Similar to the function level, the line-level performance values of ChatGPT were lower than those of C/C++, except for Top-3 Accuracy as it was not used for C/C++~\cite{fu2022linevul}.

\begin{table}[t]
\fontsize{7.5}{8.5}\selectfont
\centering
\caption{Comparisons between ChatGPT and CodeBERT for line-level SV prediction in low-resource languages.}
\begin{tabular}{lccccc}
\hline
\textbf{Lang.}        & \textbf{Model}  & \makecell[c]{\textbf{Top-3}\\ \textbf{Accuracy}} & \textbf{IFA}  & \makecell[c]{\textbf{Effort@}\\ \textbf{20\%Recall}} & \makecell[c]{\textbf{Recall@}\\ \textbf{1\%LOC}} \\ \hline
\multirow{3}{*}{Kotlin}  & GPT Fine-tuning & 87.5\%                                                           & 0.12          & 0.013                                                                & 0.114                                                            \\
                         & GPT Few-shot    & \textbf{100\%}                                                    & \textbf{0.00}    & \textbf{0.008}                                                       & 0.114                                                            \\
                         & CodeBERT (Best)  & \textbf{100\%}                                                    & \textbf{0.00}    & 0.013                                                                & \textbf{0.139}                                                   \\ \hline
\multirow{3}{*}{Swift}   & GPT Fine-tuning & 33.3\%                                                           & \textbf{4.47} & \textbf{0.016}                                                       & 0.090                                                            \\
                         & GPT Few-shot    & \textbf{43.3\%}                                                  & 11.5         & 0.024                                                                & 0.090                                                            \\
                         & CodeBERT (Best)  & 31.0\%                                                              & 9.25          & 0.024                                                                & \textbf{0.250}                                                   \\ \hline
\multirow{3}{*}{Rust}    & GPT Fine-tuning & 21.2\%                                                           & 3.29          & 0.008                                                                & 0.070                                                            \\
                         & GPT Few-shot    & \textbf{64.3\%}                                                  & \textbf{1.00}  & \textbf{0.001}                                                       & 0.037                                                            \\
                         & CodeBERT (Best)  & 58.7\%                                                           & 7.69          & 0.027                                                                & \textbf{0.216}                                                   \\ \hline \hline
\multirow{3}{*}{Avg.} & GPT Fine-tuning & 47.4\%                                                           & \textbf{2.63} & 0.013                                                                & 0.091                                                            \\
                         & GPT Few-shot    & \textbf{69.2\%}                                                  & 4.18          & \textbf{0.011}                                                       & 0.081                                                            \\
                         & CodeBERT (Best)  & 63.3\%                                                           & 5.65          & 0.022                                                                & \textbf{0.202}                                                   \\ \hline
\end{tabular}%
\label{tab:gpt-results-line-level}
\end{table}

\begin{tcolorbox}
\textbf{RQ2 Summary}.
ChatGPT performed significantly better than CodeBERT, 2.3\% -- 34.4\%$\uparrow$ at the function level and 9.3\% -- 53.5\%$\uparrow$ at the line level, for low-resource SV prediction.
ChatGPT with fine-tuning shows the best overall performance for predicting vulnerable functions. There is a performance tie between few-shot learning and fine-tuning for line-level prediction.
Despite ChatGPT's improvements, SV predictive performance, especially at the function level, in low-resource languages is still far behind that in abundant-resource languages like C/C++.
\end{tcolorbox}

\section{Threats to Validity}\label{sec:threats_to_validity}

The completeness of our datasets is a potential threat. We mitigated this by leveraging the best practice of collecting SV data from the National Vulnerability Database, the largest source of SVs in the wild, based on the methods and tools of CVEfixes~\cite{bhandari2021cvefixes}.

There are possible concerns regarding the choice and optimality of prediction models. Given resource constraints, comprehensively evaluating all available features and models becomes nearly impractical. Thus, we focused on techniques and their associated hyperparameters that have been recommended in the literature. Our pioneering work in low-resource SV prediction, despite imperfect baselines, serves as a catalyst for the evolution of more sophisticated and high-performing techniques in subsequent research.

Regarding the generalizability of our results, we only performed our study in three languages, yet we focused on the languages that are popular among developers with worldwide usage. We also used data from real-world projects of diverse domains and scales.

\section{Conclusion}
\label{sec:conclusions}

Our study addresses the challenge of SV prediction in low-resource languages. Our experiments on Kotlin, Swift, and Rust revealed that the performance of the CodeBERT-based state-of-the-art model was sub-par for function-level and line-level SV prediction in these languages.
We explored potential remedies, including data sampling techniques like random over-sampling and under-sampling, yet these approaches failed to enhance CodeBERT's performance. Intriguingly, ChatGPT showed positive results, substantially improving function-level prediction by 2.3--34.4\% and line-level prediction by 9.3--53.5\%.
While our first attempt at low-resource SV prediction for emerging yet low-resource languages is promising, there is still a long way to achieve similar performance as in abundant-resource languages like C/C++.
Our findings also underscore the pressing need for continued research in adapting and improving current SV prediction models for low-resource settings.

There are several potential future directions for SV prediction and analysis in low-resource languages. Firstly, ChatGPT can be used in conjunction with latent SVs~\cite{le2024latent} or semi-supervised learning~\cite{le2020puminer} to enhance the performance of SV prediction in low-resource languages. Secondly, besides SV prediction, future research can investigate other SV management tasks~\cite{le2019automated,le2021large,duan2021automated,le2022survey,le2022towards,fu2022vulrepair} in low-resource languages.
Thirdly, the nature of SVs in low-resource languages can change over time, so the predictions should be monitored continuously to mitigate performance degradation~\cite{arani2023sok,arani2023systematic,arani2023mitigating}.

\section*{Acknowledgments}
The work has been supported by the Cyber Security Research Centre Limited whose activities are partially funded by the Australian Government's Cooperative Research Centres Program.

\balance

\bibliographystyle{ACM-Reference-Format}
\bibliography{reference}


\begin{thebibliography}{37}


\ifx \showCODEN    \undefined \def \showCODEN     #1{\unskip}     \fi
\ifx \showDOI      \undefined \def \showDOI       #1{#1}\fi
\ifx \showISBNx    \undefined \def \showISBNx     #1{\unskip}     \fi
\ifx \showISBNxiii \undefined \def \showISBNxiii  #1{\unskip}     \fi
\ifx \showISSN     \undefined \def \showISSN      #1{\unskip}     \fi
\ifx \showLCCN     \undefined \def \showLCCN      #1{\unskip}     \fi
\ifx \shownote     \undefined \def \shownote      #1{#1}          \fi
\ifx \showarticletitle \undefined \def \showarticletitle #1{#1}   \fi
\ifx \showURL      \undefined \def \showURL       {\relax}        \fi
\providecommand\bibfield[2]{#2}
\providecommand\bibinfo[2]{#2}
\providecommand\natexlab[1]{#1}
\providecommand\showeprint[2][]{arXiv:#2}

\bibitem[\protect\citeauthoryear{Arani, Le, Zahedi, and Babar}{Arani et~al\mbox{.}}{2023a}]%
        {arani2023mitigating}
\bibfield{author}{\bibinfo{person}{Ali~Kazemi Arani}, \bibinfo{person}{Triet Huynh~Minh Le}, \bibinfo{person}{Mansooreh Zahedi}, {and} \bibinfo{person}{Muhammad~Ali Babar}.} \bibinfo{year}{2023}\natexlab{a}.
\newblock \showarticletitle{Mitigating ML Model Decay in Continuous Integration with Data Drift Detection: An Empirical Study}.
\newblock \bibinfo{journal}{\emph{arXiv preprint arXiv:2305.12736}} (\bibinfo{year}{2023}).
\newblock


\bibitem[\protect\citeauthoryear{Arani, Le, Zahedi, and Babar}{Arani et~al\mbox{.}}{2023b}]%
        {arani2023systematic}
\bibfield{author}{\bibinfo{person}{Ali~Kazemi Arani}, \bibinfo{person}{Triet Huynh~Minh Le}, \bibinfo{person}{Mansooreh Zahedi}, {and} \bibinfo{person}{Muhammad~Ali Babar}.} \bibinfo{year}{2023}\natexlab{b}.
\newblock \showarticletitle{Systematic Literature Review on Application of Machine Learning in Continuous Integration}.
\newblock \bibinfo{journal}{\emph{arXiv preprint arXiv:2305.12695}} (\bibinfo{year}{2023}).
\newblock


\bibitem[\protect\citeauthoryear{Arani, Zahedi, Le, and Babar}{Arani et~al\mbox{.}}{2023c}]%
        {arani2023sok}
\bibfield{author}{\bibinfo{person}{Ali~Kazemi Arani}, \bibinfo{person}{Mansooreh Zahedi}, \bibinfo{person}{Triet Huynh~Minh Le}, {and} \bibinfo{person}{Muhammad~Ali Babar}.} \bibinfo{year}{2023}\natexlab{c}.
\newblock \showarticletitle{Sok: Machine learning for continuous integration}. In \bibinfo{booktitle}{\emph{2023 IEEE/ACM International Workshop on Cloud Intelligence \& AIOps (AIOps)}}. IEEE, \bibinfo{pages}{8--13}.
\newblock


\bibitem[\protect\citeauthoryear{Authors}{Authors}{[n.\,d.]}]%
        {reproduction_package_ease2024}
\bibfield{author}{\bibinfo{person}{Authors}.} \bibinfo{year}{[n.\,d.]}\natexlab{}.
\newblock \bibinfo{title}{Reproduction package}.
\newblock
\newblock
\urldef\tempurl%
\url{https://github.com/lhmtriet/LLM4Vul}
\showURL{%
\tempurl}


\bibitem[\protect\citeauthoryear{Bhandari, Naseer, and Moonen}{Bhandari et~al\mbox{.}}{2021}]%
        {bhandari2021cvefixes}
\bibfield{author}{\bibinfo{person}{Guru Bhandari}, \bibinfo{person}{Amara Naseer}, {and} \bibinfo{person}{Leon Moonen}.} \bibinfo{year}{2021}\natexlab{}.
\newblock \showarticletitle{CVEfixes: automated collection of vulnerabilities and their fixes from open-source software}. In \bibinfo{booktitle}{\emph{Proceedings of the 17th International Conference on Predictive Models and Data Analytics in Software Engineering}}. \bibinfo{pages}{30--39}.
\newblock


\bibitem[\protect\citeauthoryear{Cheshkov, Zadorozhny, and Levichev}{Cheshkov et~al\mbox{.}}{2023}]%
        {cheshkov2023evaluation}
\bibfield{author}{\bibinfo{person}{Anton Cheshkov}, \bibinfo{person}{Pavel Zadorozhny}, {and} \bibinfo{person}{Rodion Levichev}.} \bibinfo{year}{2023}\natexlab{}.
\newblock \showarticletitle{Evaluation of ChatGPT Model for Vulnerability Detection}.
\newblock \bibinfo{journal}{\emph{arXiv preprint arXiv:2304.07232}} (\bibinfo{year}{2023}).
\newblock


\bibitem[\protect\citeauthoryear{Croft, Babar, and Kholoosi}{Croft et~al\mbox{.}}{2023}]%
        {croft2023data}
\bibfield{author}{\bibinfo{person}{Roland Croft}, \bibinfo{person}{M~Ali Babar}, {and} \bibinfo{person}{M~Mehdi Kholoosi}.} \bibinfo{year}{2023}\natexlab{}.
\newblock \showarticletitle{Data quality for software vulnerability datasets}. In \bibinfo{booktitle}{\emph{2023 IEEE/ACM 45th International Conference on Software Engineering (ICSE)}}. IEEE, \bibinfo{pages}{121--133}.
\newblock


\bibitem[\protect\citeauthoryear{Croft, Xie, and Babar}{Croft et~al\mbox{.}}{2022}]%
        {croft2022data}
\bibfield{author}{\bibinfo{person}{Roland Croft}, \bibinfo{person}{Yongzheng Xie}, {and} \bibinfo{person}{Muhammad~Ali Babar}.} \bibinfo{year}{2022}\natexlab{}.
\newblock \showarticletitle{Data preparation for software vulnerability prediction: A systematic literature review}.
\newblock \bibinfo{journal}{\emph{IEEE Transactions on Software Engineering}} \bibinfo{volume}{49}, \bibinfo{number}{3} (\bibinfo{year}{2022}), \bibinfo{pages}{1044--1063}.
\newblock


\bibitem[\protect\citeauthoryear{Duan, Ge, Le, Ullah, Gao, Lu, and Babar}{Duan et~al\mbox{.}}{2021}]%
        {duan2021automated}
\bibfield{author}{\bibinfo{person}{Xuanyu Duan}, \bibinfo{person}{Mengmeng Ge}, \bibinfo{person}{Triet Huynh~Minh Le}, \bibinfo{person}{Faheem Ullah}, \bibinfo{person}{Shang Gao}, \bibinfo{person}{Xuequan Lu}, {and} \bibinfo{person}{M~Ali Babar}.} \bibinfo{year}{2021}\natexlab{}.
\newblock \showarticletitle{Automated security assessment for the Internet of Things}. In \bibinfo{booktitle}{\emph{2021 IEEE 26th Pacific Rim International Symposium on Dependable Computing (PRDC)}}. IEEE, \bibinfo{pages}{47--56}.
\newblock


\bibitem[\protect\citeauthoryear{Fan, Gokkaya, Harman, Lyubarskiy, Sengupta, Yoo, and Zhang}{Fan et~al\mbox{.}}{2023}]%
        {fan2023large}
\bibfield{author}{\bibinfo{person}{Angela Fan}, \bibinfo{person}{Beliz Gokkaya}, \bibinfo{person}{Mark Harman}, \bibinfo{person}{Mitya Lyubarskiy}, \bibinfo{person}{Shubho Sengupta}, \bibinfo{person}{Shin Yoo}, {and} \bibinfo{person}{Jie~M Zhang}.} \bibinfo{year}{2023}\natexlab{}.
\newblock \showarticletitle{Large language models for software engineering: Survey and open problems}.
\newblock \bibinfo{journal}{\emph{arXiv preprint arXiv:2310.03533}} (\bibinfo{year}{2023}).
\newblock


\bibitem[\protect\citeauthoryear{Fan, Li, Wang, and Nguyen}{Fan et~al\mbox{.}}{2020}]%
        {fan2020ac}
\bibfield{author}{\bibinfo{person}{Jiahao Fan}, \bibinfo{person}{Yi Li}, \bibinfo{person}{Shaohua Wang}, {and} \bibinfo{person}{Tien~N Nguyen}.} \bibinfo{year}{2020}\natexlab{}.
\newblock \showarticletitle{A C/C++ code vulnerability dataset with code changes and CVE summaries}. In \bibinfo{booktitle}{\emph{Proceedings of the 17th International Conference on Mining Software Repositories}}. \bibinfo{pages}{508--512}.
\newblock


\bibitem[\protect\citeauthoryear{Feng, Guo, Tang, Duan, Feng, Gong, Shou, Qin, Liu, Jiang, et~al\mbox{.}}{Feng et~al\mbox{.}}{2020}]%
        {feng2020codebert}
\bibfield{author}{\bibinfo{person}{Zhangyin Feng}, \bibinfo{person}{Daya Guo}, \bibinfo{person}{Duyu Tang}, \bibinfo{person}{Nan Duan}, \bibinfo{person}{Xiaocheng Feng}, \bibinfo{person}{Ming Gong}, \bibinfo{person}{Linjun Shou}, \bibinfo{person}{Bing Qin}, \bibinfo{person}{Ting Liu}, \bibinfo{person}{Daxin Jiang}, {et~al\mbox{.}}} \bibinfo{year}{2020}\natexlab{}.
\newblock \showarticletitle{Codebert: A pre-trained model for programming and natural languages}.
\newblock \bibinfo{journal}{\emph{arXiv preprint arXiv:2002.08155}} (\bibinfo{year}{2020}).
\newblock


\bibitem[\protect\citeauthoryear{Fu and Tantithamthavorn}{Fu and Tantithamthavorn}{2022}]%
        {fu2022linevul}
\bibfield{author}{\bibinfo{person}{Michael Fu} {and} \bibinfo{person}{Chakkrit Tantithamthavorn}.} \bibinfo{year}{2022}\natexlab{}.
\newblock \showarticletitle{Linevul: A transformer-based line-level vulnerability prediction}. In \bibinfo{booktitle}{\emph{Proceedings of the 19th International Conference on Mining Software Repositories}}. \bibinfo{pages}{608--620}.
\newblock


\bibitem[\protect\citeauthoryear{Fu, Tantithamthavorn, Le, Nguyen, and Phung}{Fu et~al\mbox{.}}{2022}]%
        {fu2022vulrepair}
\bibfield{author}{\bibinfo{person}{Michael Fu}, \bibinfo{person}{Chakkrit Tantithamthavorn}, \bibinfo{person}{Trung Le}, \bibinfo{person}{Van Nguyen}, {and} \bibinfo{person}{Dinh Phung}.} \bibinfo{year}{2022}\natexlab{}.
\newblock \showarticletitle{VulRepair: A T5-based automated software vulnerability repair}. In \bibinfo{booktitle}{\emph{Proceedings of the 30th ACM Joint European Software Engineering Conference and Symposium on the Foundations of Software Engineering}}. \bibinfo{pages}{935--947}.
\newblock


\bibitem[\protect\citeauthoryear{Fu, Tantithamthavorn, Nguyen, and Le}{Fu et~al\mbox{.}}{2023}]%
        {fu2023chatgpt}
\bibfield{author}{\bibinfo{person}{Michael Fu}, \bibinfo{person}{Chakkrit Tantithamthavorn}, \bibinfo{person}{Van Nguyen}, {and} \bibinfo{person}{Trung Le}.} \bibinfo{year}{2023}\natexlab{}.
\newblock \showarticletitle{Chatgpt for vulnerability detection, classification, and repair: How far are we?}
\newblock \bibinfo{journal}{\emph{arXiv preprint arXiv:2310.09810}} (\bibinfo{year}{2023}).
\newblock


\bibitem[\protect\citeauthoryear{Hanif, Nasir, Ab~Razak, Firdaus, and Anuar}{Hanif et~al\mbox{.}}{2021}]%
        {hanif2021rise}
\bibfield{author}{\bibinfo{person}{Hazim Hanif}, \bibinfo{person}{Mohd Hairul Nizam~Md Nasir}, \bibinfo{person}{Mohd~Faizal Ab~Razak}, \bibinfo{person}{Ahmad Firdaus}, {and} \bibinfo{person}{Nor~Badrul Anuar}.} \bibinfo{year}{2021}\natexlab{}.
\newblock \showarticletitle{The rise of software vulnerability: Taxonomy of software vulnerabilities detection and machine learning approaches}.
\newblock \bibinfo{journal}{\emph{Journal of Network and Computer Applications}}  \bibinfo{volume}{179} (\bibinfo{year}{2021}), \bibinfo{pages}{103009}.
\newblock


\bibitem[\protect\citeauthoryear{Hin, Kan, Chen, and Babar}{Hin et~al\mbox{.}}{2022}]%
        {hin2022linevd}
\bibfield{author}{\bibinfo{person}{David Hin}, \bibinfo{person}{Andrey Kan}, \bibinfo{person}{Huaming Chen}, {and} \bibinfo{person}{M~Ali Babar}.} \bibinfo{year}{2022}\natexlab{}.
\newblock \showarticletitle{LineVD: Statement-level vulnerability detection using graph neural networks}. In \bibinfo{booktitle}{\emph{the 19th International Conference on Mining Software Repositories}}. \bibinfo{pages}{596--607}.
\newblock


\bibitem[\protect\citeauthoryear{Hoang, Dam, Kamei, Lo, and Ubayashi}{Hoang et~al\mbox{.}}{2019}]%
        {hoang2019deepjit}
\bibfield{author}{\bibinfo{person}{Thong Hoang}, \bibinfo{person}{Hoa~Khanh Dam}, \bibinfo{person}{Yasutaka Kamei}, \bibinfo{person}{David Lo}, {and} \bibinfo{person}{Naoyasu Ubayashi}.} \bibinfo{year}{2019}\natexlab{}.
\newblock \showarticletitle{DeepJIT: An end-to-end deep learning framework for just-in-time defect prediction}. In \bibinfo{booktitle}{\emph{2019 IEEE/ACM 16th International Conference on Mining Software Repositories (MSR)}}. IEEE, \bibinfo{pages}{34--45}.
\newblock


\bibitem[\protect\citeauthoryear{Le}{Le}{2022}]%
        {le2022towards}
\bibfield{author}{\bibinfo{person}{Triet~HM Le}.} \bibinfo{year}{2022}\natexlab{}.
\newblock \showarticletitle{Towards an improved understanding of software vulnerability assessment using data-driven approaches}.
\newblock \bibinfo{journal}{\emph{arXiv preprint arXiv:2207.11708}} (\bibinfo{year}{2022}).
\newblock


\bibitem[\protect\citeauthoryear{Le, Chen, and Babar}{Le et~al\mbox{.}}{2020a}]%
        {le2020deep}
\bibfield{author}{\bibinfo{person}{Triet~HM Le}, \bibinfo{person}{Hao Chen}, {and} \bibinfo{person}{Muhammad~Ali Babar}.} \bibinfo{year}{2020}\natexlab{a}.
\newblock \showarticletitle{Deep learning for source code modeling and generation: Models, applications, and challenges}.
\newblock \bibinfo{journal}{\emph{ACM Computing Surveys (CSUR)}} \bibinfo{volume}{53}, \bibinfo{number}{3} (\bibinfo{year}{2020}), \bibinfo{pages}{1--38}.
\newblock


\bibitem[\protect\citeauthoryear{Le, Chen, and Babar}{Le et~al\mbox{.}}{2022}]%
        {le2022survey}
\bibfield{author}{\bibinfo{person}{Triet~HM Le}, \bibinfo{person}{Huaming Chen}, {and} \bibinfo{person}{M~Ali Babar}.} \bibinfo{year}{2022}\natexlab{}.
\newblock \showarticletitle{A survey on data-driven software vulnerability assessment and prioritization}.
\newblock \bibinfo{journal}{\emph{Comput. Surveys}} \bibinfo{volume}{55}, \bibinfo{number}{5} (\bibinfo{year}{2022}), \bibinfo{pages}{1--39}.
\newblock


\bibitem[\protect\citeauthoryear{Le, Du, and Babar}{Le et~al\mbox{.}}{2024}]%
        {le2024latent}
\bibfield{author}{\bibinfo{person}{Triet~HM Le}, \bibinfo{person}{Xiaoning Du}, {and} \bibinfo{person}{M~Ali Babar}.} \bibinfo{year}{2024}\natexlab{}.
\newblock \showarticletitle{Are Latent Vulnerabilities Hidden Gems for Software Vulnerability Prediction? An Empirical Study}.
\newblock \bibinfo{journal}{\emph{arXiv preprint arXiv:2401.11105}} (\bibinfo{year}{2024}).
\newblock


\bibitem[\protect\citeauthoryear{Le and Babar}{Le and Babar}{2022}]%
        {le2022use}
\bibfield{author}{\bibinfo{person}{Triet Huynh~Minh Le} {and} \bibinfo{person}{M~Ali Babar}.} \bibinfo{year}{2022}\natexlab{}.
\newblock \showarticletitle{On the use of fine-grained vulnerable code statements for software vulnerability assessment models}. In \bibinfo{booktitle}{\emph{Proceedings of the 19th International Conference on Mining Software Repositories}}. \bibinfo{pages}{621--633}.
\newblock


\bibitem[\protect\citeauthoryear{Le, Croft, Hin, and Babar}{Le et~al\mbox{.}}{2021a}]%
        {le2021large}
\bibfield{author}{\bibinfo{person}{Triet Huynh~Minh Le}, \bibinfo{person}{Roland Croft}, \bibinfo{person}{David Hin}, {and} \bibinfo{person}{Muhammad~Ali Babar}.} \bibinfo{year}{2021}\natexlab{a}.
\newblock \showarticletitle{A large-scale study of security vulnerability support on developer q\&a websites}.
\newblock In \bibinfo{booktitle}{\emph{Evaluation and assessment in software engineering}}. \bibinfo{pages}{109--118}.
\newblock


\bibitem[\protect\citeauthoryear{Le, Hin, Croft, and Babar}{Le et~al\mbox{.}}{2020b}]%
        {le2020puminer}
\bibfield{author}{\bibinfo{person}{Triet Huynh~Minh Le}, \bibinfo{person}{David Hin}, \bibinfo{person}{Roland Croft}, {and} \bibinfo{person}{M~Ali Babar}.} \bibinfo{year}{2020}\natexlab{b}.
\newblock \showarticletitle{PUMiner: Mining security posts from developer question and answer websites with PU learning}. In \bibinfo{booktitle}{\emph{Proceedings of the 17th International Conference on Mining Software Repositories}}. \bibinfo{pages}{350--361}.
\newblock


\bibitem[\protect\citeauthoryear{Le, Hin, Croft, and Babar}{Le et~al\mbox{.}}{2021b}]%
        {le2021deepcva}
\bibfield{author}{\bibinfo{person}{Triet Huynh~Minh Le}, \bibinfo{person}{David Hin}, \bibinfo{person}{Roland Croft}, {and} \bibinfo{person}{M~Ali Babar}.} \bibinfo{year}{2021}\natexlab{b}.
\newblock \showarticletitle{Deepcva: Automated commit-level vulnerability assessment with deep multi-task learning}. In \bibinfo{booktitle}{\emph{2021 36th IEEE/ACM International Conference on Automated Software Engineering (ASE)}}. IEEE, \bibinfo{pages}{717--729}.
\newblock


\bibitem[\protect\citeauthoryear{Le, Sabir, and Babar}{Le et~al\mbox{.}}{2019}]%
        {le2019automated}
\bibfield{author}{\bibinfo{person}{Triet Huynh~Minh Le}, \bibinfo{person}{Bushra Sabir}, {and} \bibinfo{person}{Muhammad~Ali Babar}.} \bibinfo{year}{2019}\natexlab{}.
\newblock \showarticletitle{Automated software vulnerability assessment with concept drift}. In \bibinfo{booktitle}{\emph{2019 IEEE/ACM 16th International Conference on Mining Software Repositories (MSR)}}. IEEE, \bibinfo{pages}{371--382}.
\newblock


\bibitem[\protect\citeauthoryear{Li, Wang, and Nguyen}{Li et~al\mbox{.}}{2021}]%
        {li2021vulnerability}
\bibfield{author}{\bibinfo{person}{Yi Li}, \bibinfo{person}{Shaohua Wang}, {and} \bibinfo{person}{Tien~N Nguyen}.} \bibinfo{year}{2021}\natexlab{}.
\newblock \showarticletitle{Vulnerability detection with fine-grained interpretations}. In \bibinfo{booktitle}{\emph{Proceedings of the 29th ACM Joint Meeting on European Software Engineering Conference and Symposium on the Foundations of Software Engineering}}. \bibinfo{pages}{292--303}.
\newblock


\bibitem[\protect\citeauthoryear{Lin, Wen, Han, Zhang, and Xiang}{Lin et~al\mbox{.}}{2020}]%
        {lin2020software}
\bibfield{author}{\bibinfo{person}{Guanjun Lin}, \bibinfo{person}{Sheng Wen}, \bibinfo{person}{Qing-Long Han}, \bibinfo{person}{Jun Zhang}, {and} \bibinfo{person}{Yang Xiang}.} \bibinfo{year}{2020}\natexlab{}.
\newblock \showarticletitle{Software vulnerability detection using deep neural networks: A survey}.
\newblock \bibinfo{journal}{\emph{Proc. IEEE}} \bibinfo{volume}{108}, \bibinfo{number}{10} (\bibinfo{year}{2020}), \bibinfo{pages}{1825--1848}.
\newblock


\bibitem[\protect\citeauthoryear{OpenAI}{OpenAI}{[n.\,d.]}]%
        {gpt_turbo}
\bibfield{author}{\bibinfo{person}{OpenAI}.} \bibinfo{year}{[n.\,d.]}\natexlab{}.
\newblock \bibinfo{title}{OpenAI's GPT 3.5 Turbo}.
\newblock
\newblock
\urldef\tempurl%
\url{https://platform.openai.com/docs/models/gpt-3-5-turbo}
\showURL{%
\tempurl}


\bibitem[\protect\citeauthoryear{Pearce, Tan, Ahmad, Karri, and Dolan-Gavitt}{Pearce et~al\mbox{.}}{2023}]%
        {pearce2023examining}
\bibfield{author}{\bibinfo{person}{Hammond Pearce}, \bibinfo{person}{Benjamin Tan}, \bibinfo{person}{Baleegh Ahmad}, \bibinfo{person}{Ramesh Karri}, {and} \bibinfo{person}{Brendan Dolan-Gavitt}.} \bibinfo{year}{2023}\natexlab{}.
\newblock \showarticletitle{Examining zero-shot vulnerability repair with large language models}. In \bibinfo{booktitle}{\emph{2023 IEEE Symposium on Security and Privacy (SP)}}. IEEE, \bibinfo{pages}{2339--2356}.
\newblock


\bibitem[\protect\citeauthoryear{Steenhoek, Rahman, Jiles, and Le}{Steenhoek et~al\mbox{.}}{2023}]%
        {steenhoek2023empirical}
\bibfield{author}{\bibinfo{person}{Benjamin Steenhoek}, \bibinfo{person}{Md~Mahbubur Rahman}, \bibinfo{person}{Richard Jiles}, {and} \bibinfo{person}{Wei Le}.} \bibinfo{year}{2023}\natexlab{}.
\newblock \showarticletitle{An empirical study of deep learning models for vulnerability detection}. In \bibinfo{booktitle}{\emph{the 45th International Conference on Software Engineering (ICSE)}}. IEEE, \bibinfo{pages}{2237--2248}.
\newblock


\bibitem[\protect\citeauthoryear{Tomczak and Tomczak}{Tomczak and Tomczak}{2014}]%
        {tomczak2014need}
\bibfield{author}{\bibinfo{person}{Maciej Tomczak} {and} \bibinfo{person}{Ewa Tomczak}.} \bibinfo{year}{2014}\natexlab{}.
\newblock \showarticletitle{The need to report effect size estimates revisited. An overview of some recommended measures of effect size}.
\newblock \bibinfo{journal}{\emph{Trends in Sport Sciences}} \bibinfo{volume}{1}, \bibinfo{number}{21} (\bibinfo{year}{2014}), \bibinfo{pages}{19--25}.
\newblock


\bibitem[\protect\citeauthoryear{Wilcoxon}{Wilcoxon}{1992}]%
        {wilcoxon1992individual}
\bibfield{author}{\bibinfo{person}{Frank Wilcoxon}.} \bibinfo{year}{1992}\natexlab{}.
\newblock \showarticletitle{Individual comparisons by ranking methods}.
\newblock In \bibinfo{booktitle}{\emph{Breakthroughs in Statistics}}. \bibinfo{publisher}{Springer}, \bibinfo{pages}{196--202}.
\newblock


\bibitem[\protect\citeauthoryear{Zhang, Liu, Zeng, Yang, Li, and Li}{Zhang et~al\mbox{.}}{2023}]%
        {zhang2023prompt}
\bibfield{author}{\bibinfo{person}{Chenyuan Zhang}, \bibinfo{person}{Hao Liu}, \bibinfo{person}{Jiutian Zeng}, \bibinfo{person}{Kejing Yang}, \bibinfo{person}{Yuhong Li}, {and} \bibinfo{person}{Hui Li}.} \bibinfo{year}{2023}\natexlab{}.
\newblock \showarticletitle{Prompt-enhanced software vulnerability detection using chatgpt}.
\newblock \bibinfo{journal}{\emph{arXiv preprint arXiv:2308.12697}} (\bibinfo{year}{2023}).
\newblock


\bibitem[\protect\citeauthoryear{Zhou, Zhang, and Lo}{Zhou et~al\mbox{.}}{2024}]%
        {zhou2024large}
\bibfield{author}{\bibinfo{person}{Xin Zhou}, \bibinfo{person}{Ting Zhang}, {and} \bibinfo{person}{David Lo}.} \bibinfo{year}{2024}\natexlab{}.
\newblock \showarticletitle{Large language model for vulnerability detection: Emerging results and future directions}.
\newblock \bibinfo{journal}{\emph{arXiv preprint arXiv:2401.15468}} (\bibinfo{year}{2024}).
\newblock


\bibitem[\protect\citeauthoryear{Zhou, Liu, Siow, Du, and Liu}{Zhou et~al\mbox{.}}{2019}]%
        {zhou2019devign}
\bibfield{author}{\bibinfo{person}{Yaqin Zhou}, \bibinfo{person}{Shangqing Liu}, \bibinfo{person}{Jingkai Siow}, \bibinfo{person}{Xiaoning Du}, {and} \bibinfo{person}{Yang Liu}.} \bibinfo{year}{2019}\natexlab{}.
\newblock \showarticletitle{Devign: Effective vulnerability identification by learning comprehensive program semantics via graph neural networks}.
\newblock \bibinfo{journal}{\emph{Advances in neural information processing systems}}  \bibinfo{volume}{32} (\bibinfo{year}{2019}).
\newblock


\end{thebibliography}

\end{document}